# $N$=2 Structures in All String Theories

José M. Figueroa-O'Farrill[1]

*Department of Physics, Queen Mary and Westfield College
Mile End Road, London E1 4NS, UK*

The BRST cohomology of any topological conformal field theory admits the structure of a Batalin–Vilkovisky algebra, and string theories are no exception. Let us say that two topological conformal field theories are *cohomologically equivalent* if their BRST cohomologies are isomorphic as Batalin–Vilkovisky algebras. What we show in this paper is that any string theory (regardless of the matter background) is cohomologically equivalent to some twisted $N$=2 superconformal field theory. We discuss three string theories in detail: the bosonic string, the NSR string and the W$_3$ string. In each case the way the cohomological equivalence is constructed can be understood as coupling the topological conformal field theory to topological gravity. These results lend further supporting evidence to the conjecture that *any* topological conformal field theory is cohomologically equivalent to some topologically twisted $N$=2 superconformal field theory. We end the paper with some comments on different notions of equivalence for topological conformal field theories and this leads to an improved conjecture.

## 1. Introduction

Generic string theories are theories of two-dimensional quantum gravity coupled to conformal matter. Since two-dimensional gravity has no propagating degrees of freedom, it is not surprising that one can make progress in its study by studying two dimensional topological quantum field theories. The study of these theories in turn benefits from the study of those theories which in addition possess conformal invariance; since just as in the non-topological theories, topological conformal field theories (TCFTs) can be deformed to study the space of topological field theories. The study of two-dimensional TCFTs is therefore of great relevance. A large class of topological field theories can be constructed starting from any $N$=2 superconformal field theory by the twisting procedure of Witten [1] and Eguchi–Yang [2]. However one soon realises that this procedure can be generalised and that the existence of a twisted $N$=2 superconformal algebra is not a prerequisite to having a TCFT. Indeed, two other classes of TCFTs are known to exist: string theories and those obtained by twisting the Kazama algebra [3]. As shown in [4] (see also [5]) this latter class contains the TCFTs obtained from the $G/G$ gauged WZW model [6].

In [7] we conjectured, based on some preliminary investigations of the bosonic string, that in a sense which was made precise there and which will be at the heart of the present paper, both string TCFTs and Kazama TCFTs yield nothing new with respect to the class of TCFTs which can be obtained via twisting $N$=2 superconformal algebras. It is of course well-known, thanks to the works [8] and [9], that some string theories (*e.g.*, the noncritical strings, or any string with an abelian current in the matter sector) can be understood as twisted $N$=2 superconformal algebras, in the sense that in the BRST complex of the string one can embed an $N$=2 superconformal algebra. However the chiral ring of this $N$=2 superconformal algebra is generically *not* isomorphic to the BRST cohomology ring as *graded* rings. They are of course isomorphic as vector spaces and even as rings, but the gradings do not correspond, because whereas the $U(1)$ charge of the $N$=2 superconformal algebra receives a contribution from the "momentum" of the abelian current in the matter (or gravity) sector of the string theory, in the BRST cohomology the $U(1)$ charge is the ghost number. What we do in this paper is to show that given *any* string theory we can find a topologically twisted $N$=2 superconformal field theory whose chiral ring is isomorphic as a Batalin–Vilkovisky algebra (see below) to the BRST cohomology of the string theory. And this can be done regardless of the matter background of the string. Taking into account the similar result which exists for the TCFTs constructed from the Kazama algebra [10], we find that no counterexamples remain to the conjecture in [7].

The fact that all TCFTs can be obtained from the twisting procedure is not just a curiousity, but can also have important practical applications. It is very often desirable to compare TCFTs which have different descriptions: for example $G/G$ theories and noncritical W-string theories. In principle one would have to start by comparing the BRST cohomology rings, but these are not always easy to compute. We could however proceed as follows: we would first exhibit the TCFTs as twisted $N$=2 superconformal field theories, and we would then compare these instead. This would be easier since we are comparing two fixed algebraic structures before computing cohomologies. With the results of this paper, this method is actually feasible even though the conjecture remains unproven. The point is that the way by which the twisted $N$=2 superconformal field theory is constructed is uniform and (almost) algorithmic. The details may vary with each theory, but the method is the same.

This paper is organised as follows. In Section 2 we set the notation by defining what we mean by a topological conformal algebra (TCA) and review the algebraic structure inherited by the BRST cohomology of any TCA. In Section 3 we give examples of TCAs and introduce the simplest such TCA: the Koszul TCA, a twisted $N=2$ superconformal algebra which is cohomologically trivial, but which will play a very important role in the ensuing sections. In Sections 4, 5 and 6 we discuss three string theories: the bosonic string, the NSR string and the $W_3$-string respectively, and we show how to turn them into twisted $N=2$ superconformal algebras. In Section 7 we briefly talk about other string theories and in particular the $N=2$ string. In Section 8 we summarise the paper, offer some comments on the notion of equivalence in the context of topological conformal field theory and we improve the conjecture in [7].

---

[1] e-mail: j.m.figueroa@qmw.ac.uk

## 2. (Topological) Conformal Algebras

We start by reviewing the algebraic formulation of a two-dimensional topological conformal field theory. Just like conformal field theories, also topological conformal field theories enjoy holomorphic factorisation and we will restrict ourselves only to the holomorphic sector. *A conformal algebra is to a conformal field theory, as a topological conformal algebra is to a topological conformal field theory.* In this section we will not attempt to make this statement any more precise; but rather we will define what we mean by a (topological) conformal algebra. The reader is urged to use this statement either to motivate the ensuing axiomatics, or to gain an idea of what we mean by these notions. Many of the ideas in this section can be found in a variety of different conventions in the papers [11], [12], [13] and [14]. We follow most closely the formalism in the appendix to [15] or [10]. Closely related ideas but of a more geometrical flavour can be found in [12], [16], and [17].

*Conformal Algebras*

From our point of view, the definition of a conformal algebra is nothing more than an axiomatic characterisation of the operator product expansion and of the conformal properties of a two-dimensional conformal field theory. More precisely, a conformal algebra consists of the following data:

(C1) A complex vector space $\mathcal{V}$ admitting two compatible gradings: a $\mathbb{Z}_2$-grading (fermion parity) $\mathcal{V} = \mathcal{V}^{\bar{0}} \oplus \mathcal{V}^{\bar{1}}$, and a $\mathbb{Z}$-grading (conformal weight) $\mathcal{V} = \bigoplus_{n \in \mathbb{Z}} \mathcal{V}_n$. If $A \in \mathcal{V}$ is homogeneous under the $\mathbb{Z}_2$ grading, we shall denote its degree by $|A|$. Moreover the compatibility means that $\mathcal{V}$ is actually bigraded:

$$\mathcal{V} = \bigoplus_{\substack{n \in \mathbb{Z} \\ \bar{i} = \bar{0}, \bar{1}}} \mathcal{V}_n^{\bar{i}} ,$$

where $\mathcal{V}_n^{\bar{i}} = \mathcal{V}^{\bar{i}} \cap \mathcal{V}_n$.

(C2) A linear map $\mathcal{V} \to \text{End}\,\mathcal{V}[[z, z^{-1}]]$, written $A \mapsto A(z)$, which associates to every $A \in \mathcal{V}_h$ a family $\{A_n\}$ of operators in $\mathcal{V}$ defined by $A(z) = \sum_n A_n z^{-n-h}$;

(C3) A linear map $\partial : \mathcal{V}_h \to \mathcal{V}_h$ obeying $(\partial A)(z) = \frac{d}{dz} A(z)$;

(C4) An operator product expansion

$$A(z) B(w) = \sum_{n \ll \infty} \frac{[A\,B]_n(w)}{(z-w)^n} ;$$

or equivalently a family of bilinear products $[-,-]_n : \mathcal{V} \otimes \mathcal{V} \to \mathcal{V}$, for $n \in \mathbb{Z}$, such that for every $A, B \in \mathcal{V}$, $[A\,B]_n = 0$ for sufficiently large $n$, and subject to the following axioms:

- **Identity:** There exists an identity element $\mathbf{1} \in \mathcal{V}_0$ such that $\partial \mathbf{1} = 0$, and such that for all $A \in \mathcal{V}$,

$$[\mathbf{1}\,A]_n = \begin{cases} A, & \text{for } n = 0; \text{ and} \\ 0, & \text{otherwise.} \end{cases}$$

- **Commutativity:** For all $A, B \in \mathcal{V}$,

$$[B\,A]_n - (-)^{n+|A||B|}[A\,B]_n = (-)^{n+|A||B|} \sum_{\ell \geq 1} \frac{(-)^\ell}{\ell!} \partial^\ell [A\,B]_{n+\ell}$$

- **Associativity:** For all $A, B, C \in \mathcal{V}$,

$$[[A\,B]_m\,C]_n = \sum_{\ell \geq 0} (-)^\ell \binom{m-1}{\ell} \Big( [A\,[B\,C]_{n+\ell}]_{m-\ell} \\ + (-)^{m+|A||B|} [B\,[A\,C]_{\ell+1}]_{m+n-\ell-1} \Big) ,$$

where $\binom{a}{\ell} \equiv a(a-1)\cdots(a-\ell+1)/\ell!$.

(C5) An element $T \in \mathcal{V}_2$ such that for all $A \in \mathcal{V}_h$, $[T\,A]_2 = hA$ and $[T\,A]_1 = \partial A$, and such that $[T\,T]_{>4} = 0$ and $[T\,T]_4 = \frac{1}{2}c\mathbf{1}$ for some real number $c$. Notice that this implies that the field $T(z)$ generates a Virasoro algebra with central charge $c$.

The above data $(\mathcal{V}, \partial, \mathbf{1}, T, [-,-]_n)$ subject to the above axioms define what we mean by a **conformal algebra**. As useful consequences of these axioms we list the following properties, which are easy to prove:

(P1) The derivative $\partial$ is a derivation over all the brackets $[-,-]_n$, and obeys

$$[\partial A\, B]_n = (1-n)[A\, B]_{n-1} \quad \text{and} \quad [A\, \partial B]_n = \partial[A\, B]_n + (n-1)[A\, B]_{n-1}\ .$$

(P2) For $m > 0$,

$$[A\,[B\,C]_n]_m = \sum_{\ell=0}^{m-1} \binom{m-1}{\ell} [[A\,B]_{m-\ell}\,C]_{n+\ell} + (-)^{|A||B|}[B\,[A\,C]_m]_n\ .$$

Notice that for any $A$, $[A,-]_1$ is a (super)derivation over all the other brackets $[-,-]_n$.

(P3) The brackets $[-,-]_n$ have conformal weight $-n$:

$$[-,-]_n : \mathcal{V}_i \otimes \mathcal{V}_j \to \mathcal{V}_{i+j-n}\ .$$

We shall abbreviate the normal-ordered product $[A\,B]_0$ simply by $(AB)$. It follows from (P3) above that it is a graded product; but notice that it is neither associative nor commutative. In fact, it follows from the commutativity axiom that

$$(AB) - (-)^{|A||B|}(BA) = \sum_{\ell \geq 1} \frac{(-)^{1+\ell}}{\ell!} \partial^\ell [A\,B]_\ell\ ; \tag{2.1}$$

and instead of associativity, it obeys the following important rearrangement lemma:

$$(A(BC)) - (-)^{|A||B|}(B(AC)) = ((AB)C) - (-)^{|A||B|}((BA)C)\ . \tag{2.2}$$

We will often suppress the parenthesis when writing multiple normal-ordered products. The rule is that the normal-ordered product associates to the left, which means that $(ABC\cdots)$ stands for $(A(B(C\cdots)))$.

It follows from (P1) above that the $[-,-]_{\geq 0}$ imply the rest of the brackets. This explains why in practice one usually defines conformal algebras by writing down a set of generating fields and specifying the singular terms in their OPEs; equivalently the brackets $[-,-]_{>0}$. (Of course, not all such brackets are independent, since they are subject to the commutativity and associativity axioms above.) The vector space $\mathcal{V}$ is then spanned by the generating fields, their derivatives, and normal-ordered products thereof.

*Topological Conformal Algebras*

The holomorphic sector of any[2] conformal field theory has the structure of a conformal algebra as we have defined it above. Similarly, the holomorphic sector of any topological conformal field theory admits the structure of a topological conformal algebra (TCA). More precisely, a TCA is a conformal algebra which enjoys in addition the following axioms:

(T1) There exists a second integer grading (fermion number) $\mathcal{V} = \bigoplus_{n \in \mathbb{Z}} \mathcal{V}^n$ compatible with the conformal weight and extending the $\mathbb{Z}_2$-grading in the sense that the fermion parity is the reduction modulo 2 of the fermion number; that is, $\mathcal{V}^{\bar{0}} = \bigoplus_{n \in \mathbb{Z}} \mathcal{V}^{2n}$ and $\mathcal{V}^{\bar{1}} = \bigoplus_{n \in \mathbb{Z}} \mathcal{V}^{2n+1}$; and there exists an element $\mathbb{J} \in \mathcal{V}_1^0$ such that $A \in \mathcal{V}^q$ if and only if $[\mathbb{J}\,A]_1 = qA$.

(T2) There exists $\mathbb{G}^+ \in \mathcal{V}_1^1$ such that the operator $Q \equiv [\mathbb{G}^+,-]_1 : \mathcal{V} \to \mathcal{V}$ is square-zero: $Q^2 = 0$. We shall call $Q$ the BRST operator. Notice that $Q : \mathcal{V}_h^n \to \mathcal{V}_h^{n+1}$. We denote its cohomology by $H_Q^\bullet$.

(T3) $T$ has zero central charge and there exists $\mathbb{G}^- \in \mathcal{V}_2^{-1}$ such that $T = Q\mathbb{G}^-$. It is customary in a TCA to change the notation and refer to the (topological) energy-momentum tensor as $\mathbb{T}^{\mathrm{top}}$ instead of $T$. We will adhere to this convention.

In addition to these axioms, all known topological conformal field theories obey an extra postulate:

(T4) The operator $[\mathbb{G}^-,-]_2$ induces an operation $\Delta$ in BRST cohomology such that $\Delta^2 = 0$.

For the present purposes we define a **topological conformal algebra** as the data $(\mathcal{V}, \mathbb{J}, \mathbb{G}^\pm, \mathbb{T}^{\mathrm{top}})$, where $\mathcal{V}$ is a conformal algebra and where the above axioms (T1)–(T4) are obeyed.

Two remarks are in order. First of all, it is essential in order to have a nontrivial TCFT, that the topological central charge be zero. Indeed, if that were not the case, then for every $A$ such that $QA = 0$,

$$\begin{aligned}
cA &= 2[[\mathbb{T}^{\mathrm{top}}\,\mathbb{T}^{\mathrm{top}}]_4\,A]_0 & \text{by (C5)} \\
&= 2[[[\mathbb{G}^+\,\mathbb{G}^-]_1\,\mathbb{T}^{\mathrm{top}}]_4\,A]_0 & \text{by (T3)} \\
&= 2[[\mathbb{G}^+\,[\mathbb{G}^-\,\mathbb{T}^{\mathrm{top}}]_4]_1\,A]_0 & \text{using (P2)} \\
&= 2[\mathbb{G}^+\,[[\mathbb{G}^-\,\mathbb{T}^{\mathrm{top}}]_4\,A]_0]_1 & \text{using (P2) again} \\
\therefore\ A &= Q\left(\tfrac{2}{c}[[\mathbb{G}^-\,\mathbb{T}^{\mathrm{top}}]_4\,A]_0\right)\ ,
\end{aligned}$$

---

[2] without logarithmic singularities and assuming for simplicity that only fields of integer conformal weight are present. This latter restriction is of course straightforward to lift.

whence the cohomology would be empty. The second remark concerns the postulate (T4). Under some further assumptions about the generators of the TCA, it can be proven that (T4) follows from the other axioms. This would be the case, for instance, if $\mathbb{G}^-$ were a topological primary field (see below). We believe that, in fact, (T4) is superfluous; but so far we have no proof.

*Operations in BRST Cohomology*

The BRST cohomology $H_Q^\bullet$ of a topological conformal algebra inherits several algebraic operations. First of all notice that because the BRST operator acts like a derivation over all the brackets $[-,-]_n$, these descend to brackets in cohomology. Indeed, if $QA = QB = 0$, then $Q[AB]_n = 0$ for all $n$. Similarly, if in addition either $A$ or $B$ is BRST-exact, then so is $[AB]_n$ for all $n$. We will focus on the normal-ordered product, since the other brackets are trivial for a TCA. We will prove that the normal-ordered product induces a graded associative and commutative multiplication. The first thing to notice is that if $A$ is BRST-invariant, then $\partial A$ is BRST-exact:

$$\begin{aligned}
\partial A &= [\mathbb{T}^{\text{top}} A]_1 & \text{by (C5)} \\
&= [[\mathbb{G}^+ \mathbb{G}^-]_1 A]_1 & \text{by (T3)} \\
&= [\mathbb{G}^+ [\mathbb{G}^- A]_1]_1 & \text{using (P2)} \\
\therefore \quad \partial A &= Q([\mathbb{G}^- A]_1) \ . & (2.3)
\end{aligned}$$

Using this fact we can now prove that the normal-ordered product is commutative in cohomology. More concretely, if $QA = QB = 0$, then $(AB) - (-)^{|A||B|}(BA)$ is BRST-exact. Indeed, using (2.1) it follows that

$$\begin{aligned}
(AB) - (-)^{|A||B|}(BA) &= \sum_{\ell \geq 1} \frac{(-)^{1+\ell}}{\ell!} \partial^\ell [AB]_\ell \\
&= \partial \left( \sum_{\ell \geq 1} \frac{(-)^{1+\ell}}{\ell!} \partial^{\ell-1} [AB]_\ell \right) \\
&= Q \left( \sum_{\ell \geq 1} \frac{(-)^{1+\ell}}{\ell!} \partial^{\ell-1} [\mathbb{G}^- [AB]_\ell]_1 \right) \ , & (2.4)
\end{aligned}$$

where we have used (2.3) and (P1). The commutativity of the normal-ordered product actually implies its associativity. But before proving this, let us introduce a useful piece of notation. If $A$ and $B$ are BRST-invariant, we will use the expression $A \sim B$ whenever their difference is BRST-exact. Now, notice that from (2.2) and from the commutativity, the first of the following identities follows:

$$\begin{aligned}
(A(BC)) &\sim (-)^{|A||B|}(B(AC)) & (2.5) \\
&\sim (-)^{|A|(|B|+|C|)}(B(CA)) & \text{by (2.4)} \\
&\sim (-)^{|A|(|B|+|C|)+|B||C|}(C(BA)) & \text{by (2.5)} \\
&\sim (-)^{(|A|+|B|)|C|}(C(AB)) & \text{by (2.4)} \\
&\sim ((AB)C) \ . & \text{by (2.4) again}
\end{aligned}$$

That is, the normal-ordered product defines a commutative associative graded multiplication:

$$\bullet : H_Q^p \otimes H_Q^q \to H_Q^{p+q} \ .$$

This is not all the algebraic structure that the BRST cohomology inherits; the postulate (T4) lends it further structure. First let us see that the operation $\Delta$ of (T4) is well-defined in cohomology. Let $A \in \mathcal{V}_h$ be BRST-invariant; hence

$$\begin{aligned}
Q([\mathbb{G}^- A]_2) &= [\mathbb{G}^+ [\mathbb{G}^- A]_2]_1 \\
&= [\mathbb{T}^{\text{top}} A]_2 & \text{by (P2)} \\
&= hA \ , & \text{by (C5)}
\end{aligned}$$

which shows, first of all, that if $h \neq 0$, then $A$ is BRST-exact; whence all the cohomology resides in the sector of the theory with zero topological conformal weight. Therefore we can choose for any BRST cohomology class a representative BRST-invariant field with zero topological conformal weight. The above result, for $h = 0$, then shows that $[\mathbb{G}^- A]_2$ is BRST-invariant and hence defines a class in BRST cohomology. The resulting operator in cohomology taking the class of $A$ to the class of $[\mathbb{G}^- A]_2$ is called $\Delta$. The postulate (T4) says that the map $\Delta : H_Q^\bullet \to H_Q^{\bullet-1}$, obeys $\Delta^2 = 0$. It follows from the results in [**18**] that $(H_Q^\bullet, \bullet, \Delta)$ is a Batalin–Vilkovisky (BV) algebra [**11**] [**12**] [**13**] [**14**].

Let us pause to prove that (T4) follows if we take $\mathbb{G}^-$ to be a topological primary field. This proof was obtained in collaboration with Takashi Kimura; a slightly different proof was obtained independently by Füsun Akman [**14**]. Let $A$ be any BRST-invariant field. Then,

$$\begin{aligned}
\Delta^2 A &= [\mathbb{G}^- [\mathbb{G}^- A]_2]_2 \\
&= [[\mathbb{G}^- \mathbb{G}^-]_1 A]_3 + [[\mathbb{G}^- \mathbb{G}^-]_2 A]_2 - [\mathbb{G}^- [\mathbb{G}^- A]_2]_2 & \text{by (P2)} \\
\therefore \quad \Delta^2 A &= \tfrac{1}{2} [[\mathbb{G}^- \mathbb{G}^-]_1 A]_3 + \tfrac{1}{2} [[\mathbb{G}^- \mathbb{G}^-]_2 A]_2 \ .
\end{aligned}$$

But by (2.1),

$$[\mathbb{G}^-\,\mathbb{G}^-]_2 = \tfrac{1}{2}\sum_{\ell\geq 1}\frac{(-)^{\ell+1}}{\ell!}\partial^\ell[\mathbb{G}^-\,\mathbb{G}^-]_{2+\ell}\ .$$

Since (P1) implies that for any $A$ and $B$, $[\partial^2 A\,B]_2 = 0$, only the first term in the above sum contributes:

$$\begin{aligned}[][\mathbb{G}^-\,\mathbb{G}^-]_2\,A]_2 &= \tfrac{1}{2}[\partial[\mathbb{G}^-\,\mathbb{G}^-]_3\,A]_2 \\ &= -\tfrac{1}{2}[[\mathbb{G}^-\,\mathbb{G}^-]_3\,A]_1\ ;\qquad \text{using (P1) again}\end{aligned}$$

whence

$$\Delta^2 A = \tfrac{1}{2}[[\mathbb{G}^-\,\mathbb{G}^-]_1\,A]_3 - \tfrac{1}{4}[[\mathbb{G}^-\,\mathbb{G}^-]_3\,A]_1\ . \tag{2.6}$$

This is a general result, which does not depend on any conformal properties of $\mathbb{G}^-$ nor on $A$ being BRST-invariant. Now, the first thing to notice is that if $\mathbb{G}^-$ is a topological primary, then both $[\mathbb{G}^-\,\mathbb{G}^-]_1$ and $[\mathbb{G}^-\,\mathbb{G}^-]_3$ are BRST-invariant. First notice that

$$\begin{aligned}Q[\mathbb{G}^-\,\mathbb{G}^-]_3 &= [\mathbb{T}^{\mathrm{top}}\,\mathbb{G}^-]_3 - [\mathbb{G}^-\,\mathbb{T}^{\mathrm{top}}]_3 \\ &= 2[\mathbb{T}^{\mathrm{top}}\,\mathbb{G}^-]_3 + \sum_{\ell\geq 1}\frac{(-)^\ell}{\ell!}\partial^\ell[\mathbb{T}^{\mathrm{top}}\,\mathbb{G}^-]_{3+\ell}\ ,\qquad \text{using (2.1)}\end{aligned}$$

which vanishes if $\mathbb{G}^-$ is a primary, since $[\mathbb{T}^{\mathrm{top}}\,\mathbb{G}^-]_{\geq 3} = 0$. Similarly, using (2.1), (C5) and the fact that $\mathbb{G}^-$ has topological conformal weight 2, we find that

$$\begin{aligned}Q[\mathbb{G}^-\,\mathbb{G}^-]_1 &= [\mathbb{T}^{\mathrm{top}}\,\mathbb{G}^-]_1 - [\mathbb{G}^-\,\mathbb{T}^{\mathrm{top}}]_1 \\ &= \sum_{\ell\geq 2}\frac{(-)^\ell}{\ell!}\partial^\ell[\mathbb{T}^{\mathrm{top}}\,\mathbb{G}^-]_{1+\ell}\ ,\end{aligned}$$

which is again zero for $\mathbb{G}^-$ a topological primary. But now $[\mathbb{G}^-\,\mathbb{G}^-]_1$ and $[\mathbb{G}^-\,\mathbb{G}^-]_3$ are BRST-invariant fields with nonzero topological conformal weights (3 and 1, respectively), hence by the previous discussion they are BRST-exact. Since $A$ is BRST-invariant, then both $[[\mathbb{G}^-\,\mathbb{G}^-]_3\,A]_1$ and $[[\mathbb{G}^-\,\mathbb{G}^-]_1\,A]_3$ are BRST-exact; and hence so is their sum. This proves that $\Delta^2 = 0$ in cohomology, and hence (T4). Notice that this result is in some sense stronger than what was needed, since it is only the RHS of (2.6) that need be BRST-exact and not each term separately. This probably means that $\mathbb{G}^-$ need not be assumed primary for (T4) to hold.

Let us end this section with the following definition. We say that two TCAs are **cohomologically equivalent** if and only if their BRST cohomologies are isomorphic as BV algebras. The extent to which cohomological equivalence characterises the topological conformal field theory described by the TCA will be briefly discussed in the concluding section.

## 3. Some Examples

In this section we look at some examples of TCAs. There are three main categories of examples: the twisted $N=2$ superconformal algebras, the TCAs arising in string theory, and the TCA constructed by Kazama in [**3**].

*Twisted $N=2$ Superconformal Algebras*

The simplest example of a TCA, as the notation suggests, is the twisted $N=2$ superconformal algebra, which is generated by the fields $\mathbb{J}$, $\mathbb{G}^\pm$, and $\mathbb{T}^{\mathrm{top}}$ subject to the following operator product expansions:

$$\begin{aligned}\mathbb{G}^\pm(z)\mathbb{G}^\pm(w) &= \mathrm{reg} \\ \mathbb{G}^+(z)\mathbb{G}^-(w) &= \frac{d}{(z-w)^3} + \frac{\mathbb{J}(w)}{(z-w)^2} + \frac{\mathbb{T}^{\mathrm{top}}(w)}{z-w} + \mathrm{reg} \\ \mathbb{J}(z)\mathbb{G}^\pm(w) &= \frac{\pm\mathbb{G}^\pm(w)}{z-w} + \mathrm{reg}\end{aligned} \tag{3.1}$$

These are not all the nontrivial OPEs in the twisted $N=2$ superconformal algebra, but the others follow from these [**19**] [**15**]. In this case the BRST cohomology is known as the chiral ring. Different realisations of the twisted $N=2$ superconformal algebra will give rise to different chiral rings.

The simplest nontrivial chiral ring comes from the following realisation. We take our conformal algebra to be the one generated by two BC systems: one fermionic $(b,c)$ and one bosonic $(\beta,\gamma)$, subject to the usual OPEs:

$$b(z)c(w) = \frac{1}{z-w}\quad\text{and}\quad \beta(z)\gamma(w) = \frac{1}{z-w}\ . \tag{3.2}$$

The TCA is defined by the following fields

$$\begin{aligned}\mathbb{G}^+_{\mathcal{K}} &= b\gamma \\ \mathbb{G}^-_{\mathcal{K}} &= \lambda\partial c\beta + (\lambda-1)c\partial\beta \\ \mathbb{J}_{\mathcal{K}} &= (1-\lambda)bc + \lambda\beta\gamma \\ \mathbb{T}^{\mathrm{top}}_{\mathcal{K}} &= \lambda\left(\beta\partial\gamma - b\partial c\right) + (\lambda-1)\left(\partial\beta\gamma - \partial bc\right)\ ,\end{aligned} \tag{3.3}$$

which, for any $\lambda$, satisfy a topologically twisted $N=2$ superconformal algebra. When $\lambda = 2$, there is a deformation of this realisation which consists of adding to $\mathbb{G}^-_{\mathcal{K}}$ a term $\mu b$, for any $\mu$. For $\lambda = 2$ the Koszul TCA is also the semi-infinite Weil complex of the Virasoro algebra and plays a crucial role in the coupling of topological conformal field theories to topological gravity [**12**].

All these realisations (for any $\lambda$ and $\mu$) share the same chiral ring, since in fact the BRST operator remains unmodified. It is not hard to prove (using, for example, the Kugo-Ojima mechanism) that the chiral ring of this twisted $N$=2 superconformal algebra is actually

$$H_Q^n \cong \begin{cases} \mathbb{C}\mathbf{1}, & n=0; \text{ and} \\ 0, & \text{otherwise.} \end{cases}$$

Hence it is in some sense the simplest example of a topological conformal field theory. We will call this TCA a **Koszul** TCA.

Notice that if $\mathcal{T} = (\mathcal{V}, \mathbb{J}, \mathbb{G}^\pm, \mathbb{T}^{\text{top}})$ and $\mathcal{T}' = (\mathcal{V}', \mathbb{J}', \mathbb{G}^{\pm\prime}, \mathbb{T}^{\text{top}\prime})$ are two TCAs, so is their tensor product $\mathcal{T} \otimes \mathcal{T}' = (\mathcal{V} \otimes \mathcal{V}', \mathbb{J}+\mathbb{J}', \mathbb{G}^\pm+\mathbb{G}^{\pm\prime}, \mathbb{T}^{\text{top}}+\mathbb{T}^{\text{top}\prime})$. Since the BRST charge of the tensor product theory is the sum of the BRST charges, the Künneth theorem says that the BRST cohomology of the tensor product theory is the tensor product of the BRST cohomologies. Furthermore it is not hard to prove that the isomorphism is one of BV algebras. In particular, if $\mathcal{T}$ is any TCA and $\mathcal{K}$ is a Koszul TCA, then $\mathcal{T}$ is cohomologically equivalent to $\mathcal{T} \otimes \mathcal{K}$. This fact will be very useful in what follows.

*The Kazama TCA*

Not all TCAs are twisted $N$=2 superconformal algebras, however. As counterexamples we may state the TCA constructed by Kazama in [**3**] and the TCAs appearing in many string theories. We shall have more to say about the string theories in the sections to come, so in this subsection we will only offer some comments on the Kazama TCA.

The search for TCAs which would generalise the twisted $N$=2 superconformal algebra, led Kazama [**3**] to the following coformal algebra. It is generated by fields $\mathbb{T}^{\text{top}}$, $\mathbb{G}^\pm$, $\mathbb{J}$, $\Phi$, and $\mathbb{F}$ subject to the following OPEs:

$$\mathbb{G}^+(z)\mathbb{G}^+(w) = \text{reg}$$
$$\mathbb{G}^+(z)\mathbb{G}^-(w) = \frac{d}{(z-w)^3} + \frac{\mathbb{J}(w)}{(z-w)^2} + \frac{\mathbb{T}^{\text{top}}(w)}{z-w} + \text{reg}$$
$$\mathbb{J}(z)\mathbb{G}^\pm(w) = \frac{\pm\mathbb{G}^\pm(w)}{z-w} + \text{reg}$$
$$\mathbb{G}^-(z)\mathbb{G}^-(w) = \frac{-2\mathbb{F}(w)}{z-w} + \text{reg}$$
$$\mathbb{G}^+(z)\Phi(w) = \frac{\mathbb{F}(w)}{z-w} + \text{reg}$$
$$\mathbb{J}(z)\Phi(w) = \frac{-3\Phi(w)}{z-w} + \text{reg} \ .$$

As in the twisted $N$=2 superconformal algebra, there are more nonzero OPEs but they are uniquely characterised by these [**10**]. The Kazama algebra appears naturally in the context of the $G/G$ gauged WZW model [**4**] [**5**] and, more generally, there exists a construction in terms of Manin pairs [**10**]. Notice that if we put $\Phi(z) = \mathbb{F}(z) = 0$, then the Kazama algebra reduces to an $N$=2 superconformal algebra; but even for nonvanishing $\Phi(z)$ and $\mathbb{F}(z)$, the Kazama algebra is actually a topological conformal algebra. To see this it is enough to notice that (T1)–(T3) are manifestly true; and since (as can be deduced from the above OPEs) $\mathbb{G}^-$ is a topological primary, by the discussion at the end of the last section, (T4) follows.

Nevertheless, despite the fact that in general, the Kazama TCA is not a twisted $N$=2 superconformal algebra, it follows from the results of Getzler in [**10**] that it is *cohomologically equivalent* to one. Indeed, tensoring the Kazama TCA with a $(\lambda = 2, \mu = 1)$ Koszul TCA, one can modify the fields $(\mathbb{J}, \mathbb{G}^\pm, \mathbb{T}^{\text{top}})$ without changing the BV algebra structure in BRST cohomology, in such a way that they obey (3.1). The details can be found in [**10**].

*Some prefatory remarks about string theories*

What about string theories? Are they also cohomologically equivalent to $N$=2 superconformal field theories? It has been known for some time [**8**], that the TCA arising from noncritical bosonic string theories can be modified to a topologically twisted $N$=2 superconformal algebra. Similar results for most (but not all) string theories appeared in [**9**]. All these results, however share one thing in common. Whereas that BRST charge in the string TCA and in the $N$=2 superconformal algebra agree, so that the cohomologies are isomorphic; the cohomologies are graded differently in both cases. The reason is that in the string description the cohomology is graded by ghost number, whereas in the twisted $N$=2 description the cohomology is graded by the $U(1)$ charge which receives contributions from the abelian current in the gravity or matter sector of the string. For the noncritical string theory, for instance, the grading of the cohomology in the $N$=2 description is a linear combination of the Liouville momentum and the ghost number. Thus the results in [**8**] and [**9**] do not provide us with cohomological equivalences.

In [**7**] we used the embedding of the bosonic string into the NSR string [**20**] [**21**] [**22**] to prove that *any* bosonic string theory is cohomologically equivalent to a twisted $N$=2 superconformal field theory. Unravelling the construction revealed a general method of constructing cohomological equivalences which makes no reference to string embeddings and which can be understood as coupling to topological gravity; although this is not essential. We therefore conjectured that all TCAs are cohomologically equivalent to some twisted $N$=2 superconformal algebra. The rest of the paper is devoted to providing ample evidence in support of this conjecture. We will consider in detail three string theories: the bosonic string, the NSR string and the $W_3$-string. The strategy in all cases will be the same. We will tensor the string TCA with a Koszul TCA— which we saw to be a cohomological equivalence—and we will then deform the generators in the tensor product theory to satisfy (3.1). In all cases, the construction works for any value of the weight $\lambda$ of the Koszul TCA; but when we take $\lambda = 2$ we can interpret the construction as coupling to topological gravity. (Of course, when actually computing the spectrum of the coupled theory we would have to compute the equivariant BRST cohomology [**12**].)

## 4. The Bosonic String

In this section we discuss the topological conformal algebra defined by the bosonic string and we describe how to turn it into a twisted $N$=2 superconformal algebra by tensoring it with a Koszul TCA, as described in the previous section.

Matters of manifest spacetime interpretation aside, any conformal field theory with central charge $c_M$=26 is a consistent bosonic string background. To quantise the theory we introduce fermionic ghosts $(\tilde{b}, \tilde{c})$ of weights $(2, -1)$ and we define the following composite fields:

$$\begin{aligned}
\mathbb{G}^+_{N=0} &= T_M \tilde{c} + \tilde{b}\tilde{c}\partial\tilde{c} + \tfrac{3}{2}\partial^2 \tilde{c} \\
\mathbb{G}^-_{N=0} &= \tilde{b} \\
\mathbb{J}_{N=0} &= -\tilde{b}\tilde{c} \\
\mathbb{T}^{\text{top}}_{N=0} &= T_M - 2\tilde{b}\partial\tilde{c} - \partial\tilde{b}\tilde{c} \ ,
\end{aligned} \quad (4.1)$$

where $T_M$ is the energy-momentum tensor describing the string background. The above fields generate the topological conformal algebra of the bosonic string, which closes upon adding two more fields $\tilde{c}$ and $\tilde{c}\partial\tilde{c}$. It suffices to notice that the OPE of the BRST current $\mathbb{G}^+_{N=0}$ with itself is not regular, to conclude that this topological conformal algebra is not a twisted $N$=2 superconformal algebra. Since the BRST current is defined only up to a total derivative, one may try to "improve" it to cancel the singular part of the OPE; but it is easily shown that this is impossible with only the fields that we have available: $T_M$, $\tilde{c}$, and $\tilde{b}$; that is, for generic background. In fact, we have already added an improvement term to the naive BRST current in order to cancel the first order pole (equivalently, to make it a primary field). Of course, as is well-known, for special backgrounds there may be ways to improve the BRST current and the $U(1)$ current $\mathbb{J}_{N=0}$ to make a twisted $N$=2 superconformal algebra. This is the case, for example, for noncritical strings [**8**] [**9**], where $T_M$ has a Liouville part.

In [**7**] we showed how embedding the $N$=0 string into the NSR string, one can actually improve the above fields to make them obey (3.1) and hence define an honest $N$=2 superconformal algebra. We also showed how these fields could be conjugated via an automorphism of the underlying conformal algebra to fields whose forms suggest tensoring with a Koszul TCA. Indeed, the conjugated fields are given by[3]:

$$\begin{aligned}
\mathbb{G}^+ &= \mathbb{G}^+_{N=0} + \mathbb{G}^+_{\mathcal{K}} - \partial X \\
\mathbb{G}^- &= \mathbb{G}^-_{N=0} + \mathbb{G}^-_{\mathcal{K}} \\
\mathbb{J} &= \mathbb{J}_{N=0} + \mathbb{J}_{\mathcal{K}} + (bc - \beta\gamma) + \partial\left(\tilde{c}c\beta\right) \\
\mathbb{T}^{\text{top}} &= \mathbb{T}^{\text{top}}_{N=0} + \mathbb{T}^{\text{top}}_{\mathcal{K}} \ ,
\end{aligned} \quad (4.2)$$

where $X$ is given by

$$X = \tilde{c}(bc - \beta\gamma) + \beta c \tilde{c}\partial\tilde{c} \ ,$$

and where the fields generating the Koszul TCA algebra can be read off from (3.3). Notice that the value of $\lambda$ in (3.3) is still arbitrary.

Notice that since $\mathbb{G}^+$ only gets deformed by a total derivative, the BRST charge is the same as in the tensor product $(N$=$0) \otimes \mathcal{K}$. Similarly all the other algebraic structures remain as in the tensor product $(N$=$0) \otimes \mathcal{K}$, except for the $U(1)$ charge which receives a correction in the Koszul sector. However this sector is cohomologically trivial, and so this correction is invisible in cohomology. But now the tensor product $(N$=$0) \otimes \mathcal{K}$ is cohomologically equivalent to the TCA of the $N$=0 string. Hence we conclude that the topological conformal algebra (4.1) of any bosonic string is cohomologically equivalent to the above twisted $N$=2 superconformal algebra (4.2).

## 5. The NSR String

In this section we prove that the topological conformal algebra defined by the NSR string is cohomologically equivalent to a twisted $N$=2 superconformal algebra. This fact has already been established by Marcus in [**24**] by untwisting the embedding of the NSR string into an $N$=2 string [**20**]. This result depends crucially on the bosonisation of the superconformal ghosts. In this section we prove this simply by tensoring with a Koszul TCA, just like we did for the $N$=0 string in the last section. This avoids having to bosonise anything, but does of course introduce new fields. However this is in step with the general method to construct cohomological equivalences advocated in this paper and in [**7**].

Any $N$=1 superconformal algebra generated by $(T_M, G_M)$ with $c_M = 15$ is a consistent background for the NSR string. In order to describe the BRST complex of the NSR string, we introduce two ghost systems: one fermionic $(\tilde{b}, \tilde{c})$ and one bosonic $(\tilde{\beta}, \tilde{\gamma})$, with OPEs given *mutatis mutandis* by (3.2). The fermionic BC system has weights $(2, -1)$, whereas the bosonic BC system has weights $(\frac{3}{2}, -\frac{1}{2})$. The topological conformal algebra of the NSR string is defined by the following fields:

$$\begin{aligned}
\mathbb{G}^+_{N=1} &= T_M \tilde{c} + G_M \tilde{\gamma} + \tilde{b}\tilde{c}\partial\tilde{c} - \tilde{b}\tilde{\gamma}^2 + \tilde{\beta}\tilde{\gamma}\tilde{c} - \tfrac{1}{2}\tilde{\beta}\tilde{\gamma}\partial\tilde{c} + \tfrac{1}{2}\partial^2\tilde{c} - \tfrac{1}{2}\partial\left(\tilde{\beta}\tilde{\gamma}\tilde{c}\right) \\
\mathbb{G}^-_{N=1} &= \tilde{b} \\
\mathbb{J}_{N=1} &= -\tilde{b}\tilde{c} + \tilde{\beta}\tilde{\gamma} \\
\mathbb{T}^{\text{top}}_{N=1} &= T_M - 2\tilde{b}\partial\tilde{c} - \partial\tilde{b}\tilde{c} + \tfrac{3}{2}\tilde{\beta}\partial\tilde{\gamma} + \tfrac{1}{2}\partial\tilde{\beta}\tilde{\gamma} ,
\end{aligned} \quad (5.1)$$

where as in the bosonic string we have already improved the BRST current to make it a primary field. As in the bosonic string, for generic background $(T_M, G_M)$ it is impossible to improve the above fields without modifying the BRST cohomology in such a way that $(\mathbb{J}_{N=1}, \mathbb{G}^\pm{}_{N=1}, \mathbb{T}^{\text{top}}{}_{N=1})$ generate a twisted $N=2$ superconformal algebra—at least without employing nonstandard bosonisation techniques as in [24]. Of course, for some backgrounds (including noncritical $N=1$ strings) this is of course possible [9].

---

[3] Needless to say, most of the explicit calculations in this paper have been either done or checked with *OPEdefs* [**23**], the wonderful *Mathematica* package written and continuously improved by Kris Thielemans.

We will now show that the topological conformal algebra defined by (5.1) is cohomologically equivalent to a twisted $N=2$ superconformal algebra. We first tensor by a Koszul TCA with arbitrary $\lambda$, and we then deform the fields in a suitable fashion. The following fields

$$\begin{aligned}
\mathbb{G}^+ &= \mathbb{G}^+_{N=1} + \mathbb{G}^+_{\mathcal{K}} - \partial Y \\
\mathbb{G}^- &= \mathbb{G}^-_{N=1} + \mathbb{G}^-_{\mathcal{K}} \\
\mathbb{J} &= \mathbb{J}_{N=1} + \mathbb{J}_{\mathcal{K}} + \tfrac{1}{2}(bc - \beta\gamma) + \tfrac{1}{2}\partial\left(\tilde{c}c\beta\right) \\
\mathbb{T}^{\text{top}} &= \mathbb{T}^{\text{top}}_{N=1} + \mathbb{T}^{\text{top}}_{\mathcal{K}} ,
\end{aligned}$$

where $Y$ is given by

$$Y = \tilde{c}(bc - \beta\gamma) + \beta c(\tilde{c}\partial\tilde{c} - \tilde{\gamma}^2) ,$$

can be shown to satisfy the defining OPEs (3.1) for a twisted $N=2$ superconformal algebra.

As in the previous section, notice that all the fields are essentially as in the tensor product $(N=1) \otimes \mathcal{K}$, except for the $U(1)$ charge assignments. But this difference does not affect the cohomology since it involves only the cohomologically trivial Koszul sector. Since the tensor product $(N=1) \otimes \mathcal{K}$ is cohomologically equivalent to the $N=1$ TCA, we have shown that any NSR string is cohomologically equivalent to a twisted $N=2$ superconformal algebra.

## 6. The $\mathsf{W}_3$-String

As our final example of a string theory, we discuss the BRST complex of the (critical) $\mathsf{W}_3$-string. Given any realisation $(T_M, W_M)$ of the $\mathsf{W}_3$ algebra with $c_M = 100$, we can construct a consistent $\mathsf{W}_3$-string theory and therefore an associated topological conformal algebra. This algebra is embedded in the conformal algebra generated by the fields $(T_M, W_M, \tilde{b}_1, \tilde{c}_1, \tilde{b}_2, \tilde{c}_2)$, where $(\tilde{b}_i, \tilde{c}_i)$ are fermionic BC systems of conformal weights $(\lambda_i, 1-\lambda_i)$, with $\lambda_1 = 2$ and $\lambda_2 = 3$. They obey the usual OPE for fermionic BC systems (see the first OPE in (3.2)). On the other hand, $(T_M, W_M)$ obey the $\mathsf{W}_3$ algebra with $c_M = 100$. We refrain from writing it down explicitly. The generators of the topological conformal algebra of the $\mathsf{W}_3$-string are given by:

$$\begin{aligned}
\mathbb{G}^+_{\mathsf{W}_3} &= T_M \tilde{c}_1 + W_M \tilde{c}_2 + \tilde{b}_1\tilde{c}_1\partial\tilde{c}_1 - 3\tilde{c}_1\tilde{b}_2\partial\tilde{c}_2 - 2\tilde{c}_1\partial\tilde{b}_2\tilde{c}_2 + \tfrac{8}{261}T_M\tilde{b}_1\tilde{c}_2\partial\tilde{c}_2 \\
&\quad + \tfrac{25}{522}\partial\tilde{b}_1\tilde{c}_2\partial^2\tilde{c}_2 + \tfrac{125}{1566}\tilde{b}_1\tilde{c}_2\partial^3\tilde{c}_2 + \partial U \\
\mathbb{G}^-_{\mathsf{W}_3} &= \tilde{b}_1 \\
\mathbb{J}_{\mathsf{W}_3} &= -\tilde{b}_1\tilde{c}_1 - \tilde{b}_2\tilde{c}_2 \\
\mathbb{T}^{\text{top}}_{\mathsf{W}_3} &= T_M - 2\tilde{b}_1\partial\tilde{c}_1 - \partial\tilde{b}_1\tilde{c}_1 - 3\tilde{b}_2\partial\tilde{c}_2 - 2\partial\tilde{b}_2\tilde{c}_2 ,
\end{aligned}$$

where $U$ is given by

$$U = \tilde{c}_1\tilde{b}_2\tilde{c}_2 + \tfrac{25}{174}\partial\tilde{b}_1\tilde{c}_2\partial\tilde{c}_2 + \tfrac{25}{522}\tilde{b}_1\tilde{c}_2\partial^2\tilde{c}_2 ,$$

and has been so chosen to make $[\mathbb{G}^+_{\mathsf{W}_3}, \mathbb{G}^+_{\mathsf{W}_3}]_1 = 0$ or equivalently to make $\mathbb{G}^+_{\mathsf{W}_3}$ a topological primary field.

As in the other string theories analysed in the previous sections, the above TCA cannot be improved (while keeping the BRST cohomology intact) in such a way that it becomes a twisted $N=2$ superconformal algebra. Nevertheless this becomes possible after tensoring it with a Koszul TCA (for any $\lambda$). Indeed, let us define the following fields

$$\begin{aligned}
\mathbb{G}^+ &= \mathbb{G}^+_{\mathsf{W}_3} + \mathbb{G}^+_{\mathcal{K}} - \partial V \\
\mathbb{G}^- &= \mathbb{G}^-_{\mathsf{W}_3} + \mathbb{G}^-_{\mathcal{K}} \\
\mathbb{J} &= \mathbb{J}_{\mathsf{W}_3} + \mathbb{J}_{\mathcal{K}} + 3(bc - \beta\gamma) + \partial Z \\
\mathbb{T}^{\text{top}} &= \mathbb{T}^{\text{top}}_{\mathsf{W}_3} + \mathbb{T}^{\text{top}}_{\mathcal{K}} ,
\end{aligned}$$

where $V$ and $Z$ are given by

$$V = (3\tilde{c}_1 + \tfrac{25}{261}\tilde{b}_1\tilde{c}_2\partial\tilde{c}_2)(bc - \beta\gamma) + \tfrac{25}{87}\tilde{c}_2\partial\tilde{c}_2\beta\beta(c\partial\gamma - \partial c\gamma)$$

$$- \left(3\partial\tilde{c}_1\tilde{c}_1 - \tfrac{49}{261}T\tilde{c}_2\partial\tilde{c}_2 + \tfrac{175}{522}\partial\tilde{c}_2\partial^2\tilde{c}_2 - \tfrac{275}{1566}\tilde{c}_2\partial^3\tilde{c}_2\right)c\beta$$

$$- \tfrac{25}{261}\left(\partial(\tilde{b}_1\tilde{c}_1\tilde{c}_2\partial\tilde{c}_2) - 2\tilde{b}_1\partial\tilde{c}_1\tilde{c}_2\partial\tilde{c}_2\right)c\beta + \tfrac{50}{87}\tilde{c}_2\partial\tilde{c}_2 bc\partial c\beta$$

$$+ \tfrac{25}{87}\left(3\partial\tilde{c}_1\tilde{c}_2\partial\tilde{c}_2 - \tilde{c}_1\tilde{c}_2\partial^2\tilde{c}_2\right)c\partial c\beta\beta$$

and

$$Z = 3\tilde{c}_1 c\beta + \tfrac{25}{261}\tilde{b}_1\tilde{c}_2\partial\tilde{c}_2 c\beta - \tfrac{25}{87}\beta^2 c\partial c\tilde{c}_2\partial\tilde{c}_2$$

and where the Koszul fields can be read off from (3.3), where $\lambda$ is still arbitrary.

One can then prove (after some tedious calculation, even with the computer) that the above fields obey (3.1), whence they obey a twisted $N=2$ superconformal algebra. The same arguments as for the $N=0$ and NSR strings imply that this twisted $N=2$ superconformal algebra is cohomologically equivalent to the tensor product $(\mathsf{W}_3) \otimes \mathcal{K}$, which as shown in Section 3 is itself cohomologically equivalent to the $(\mathsf{W}_3)$ TCA. Hence we conclude that *any* $\mathsf{W}_3$-string is cohomologically equivalent to some twisted $N=2$ superconformal algebra.

## 7. Other String Theories

In the previous sections we have discussed what could be considered the three most representative string theories, but these are not all the string theories in existence. There are string theories based on higher W-algebras, *e.g.*, $\mathsf{W}_4$, $\mathsf{W}_{2,s}$; and also strings with $N>1$ superconformal symmetry, *e.g.*, the $N=2$ string. How about these string theories? Are they also cohomologically equivalent to twisted $N=2$ superconformal field theories?

There is little or no reason to expect that the situation for other W-strings be any different than for the $\mathsf{W}_3$-string; only that the precise details of the equivalence are bound to be messier as the complexity of the algebra increases. But for superstrings with extended $N>1$ supersymmetry, the situation is actually much simpler than for the strings discussed in this paper. Indeed, these string theories are cohomologically equivalent to a twisted $N=2$ superconformal algebra *without* the need for new fields [**25**] [**26**].

As a convincing example, let us briefly discuss the $N=2$ string. These results were obtained in [**25**]. We will let $(J_M, G_M^\pm, T_M)$ denote a $c=6$ realisation of the $N=2$ superconformal algebra – it provides a consistent background for the $N=2$ string. In order to define the theory we introduce the relevant ghosts systems: two fermionic BC systems $(b_1, c_1)$ and $(b_2, c_2)$, and two bosonic BC systems $(\beta^+, \gamma_+)$ and $(\beta^-, \gamma_-)$ of weights $(1,0)$, $(2,-1)$, $(\tfrac{3}{2}, -\tfrac{1}{2})$, and $(\tfrac{3}{2}, -\tfrac{1}{2})$, respectively. The following fields generate the topological conformal algebra of the $N=2$ string:

$$\mathbb{G}_{N=2}^+ = c_1 J_M + c_2 T_M + \gamma_+ G_M^+ + \gamma_- G_M^- + b_2 c_2 \partial c_2 - c_1(\beta^+\gamma_+ - \beta^-\gamma_-)$$
$$+ c_2(-b_1\partial c_1 + \tfrac{3}{2}\beta^+\partial\gamma_+ + \tfrac{1}{2}\partial\beta^+\gamma_+ \tfrac{3}{2}\beta^-\partial\gamma_- + \tfrac{1}{2}\partial\beta^-\gamma_-) - b_2\gamma_+\gamma_-$$
$$+ \tfrac{1}{2}b_1(\gamma_+\partial\gamma_- - \partial\gamma_+\gamma_-)$$

$$\mathbb{G}_{N=2}^- = b_2$$

$$\mathbb{J}_{N=2} = -b_1 c_1 - b_2 c_2 + \beta^+\gamma_+ + \beta^-\gamma_-$$

$$\mathbb{T}_{N=2}^{\text{top}} = T_M - b_1\partial c_1 - 2b_2\partial c_2 - \partial b_2 c_2 + \tfrac{3}{2}\beta^+\partial\gamma_+ + \tfrac{1}{2}\partial\beta^+\gamma_+$$
$$+ \tfrac{3}{2}\beta^-\partial\gamma_- + \tfrac{1}{2}\partial\beta^-\gamma_- \ .$$

These fields don't quite obey (3.1); but one can easily find a deformation of these currents which does. In fact, all we have to do is to add to $\mathbb{G}_{N=2}^+$ a term $-\partial W$, where

$$W = c_2 \mathbb{J}_{N=2} = c_2(-b_1 c_1 + \beta^+\gamma_+ + \beta^-\gamma_-) \ . \tag{7.1}$$

Notice that the improved fields obey (3.1) and since all we have done is add a total derivative to the BRST current, the cohomology remains unchanged, even as a BV algebra. Hence the $N=2$ string is cohomologically equivalent to a twisted $N=2$ superconformal algebra. Notice that the untwisted $N=2$ superconformal algebra has zero central charge as well; that is, $\mathbb{J}_{N=2}$ is a null current. This was the crucial observation in [**25**].

Let us end this section with a curious fact. A little more work reveals that there is a one parameter family of such twisted $N=2$ superconformal algebras. Indeed, let us define the following fields:

$$\mathbb{G}^+ = \mathbb{G}_{N=2}^+ - \partial W + \lambda\partial\left(c_2(J_M - \beta^+\gamma_+ + \beta^-\gamma_-) + b_1\gamma_+\gamma_-\right)$$
$$\mathbb{G}^- = \mathbb{G}_{N=2}^-$$
$$\mathbb{J} = \mathbb{J}_{N=2} - \lambda\left(J_M - \beta^+\gamma_+ + \beta^-\gamma_-\right)$$
$$\mathbb{T}^{\text{top}} = \mathbb{T}_{N=2}^{\text{top}} \ ,$$

where $W$ is still given by (7.1). These fields obey (3.1) with zero (untwisted) central charge. However we cannot claim that for all values of the parameter $\lambda$ we have a cohomological equivalence with the $N=2$ string. The reason is that the cohomology is graded differently for different values of $\lambda$, and only for $\lambda = 0$ does the grading agree with the original one by ghost number.

# 8. Summary and an Improved Conjecture

To summarise, we have shown what we believe to be incontrovertible evidence that any string theory is cohomologically equivalent to some topological conformal field theory obtained by twisting an underlying $N{=}2$ superconformal algebra; that is, that given any string theory there exists some twisted $N{=}2$ superconformal field theory whose chiral ring coincides with the physical spectrum of the string, not just as a (graded) vector space but indeed as a Batalin–Vilkovisky algebra. We have seen this explicitly for the $N{\leq}2$ and $\mathsf{W}_3$ strings, but there is no doubt in our mind that this is true in general. Together with the similar result [10] of Getzler's for the Kazama topological conformal algebra, we are left with no counterexamples to the conjecture [7] that all TCFTs are cohomologically equivalent to (topologically twisted) $N{=}2$ superconformal field theories. The crucial point is to notice that tensoring a given topological conformal algebra with the Koszul TCA of (3.3) is a cohomological equivalence. This then gives us sufficient freedom to modify the generators to make them obey the OPEs (3.1) of a topologically twisted $N{=}2$ superconformal algebra. Notice that this method can be interpreted as coupling to topological gravity. In this sense, this situation is very reminiscent to the fact that any generally covariant theory in two-dimensions may be coupled to Liouville theory in order to obtain a conformal invariant theory. In this sense, the Koszul TCA is to $N{=}2$ supercofonformal symmetry what Liouville theory is to conformal symmetry.

*How much of an equivalence is "cohomological equivalence"?*

The answer to this question clearly depends on how much of the structure of a topological conformal field theory is characterised by the Batalin–Vilkovisky algebra structure in cohomology. A simple topological analogy might prove useful. In classical topology one may ask the following question: *To what extent is a manifold characterised by its de Rham cohomology ring?* One might hope that the de Rham cohomology ring would characterise the manifold topologically, but it is easy to see that it is a weaker invariant, since it is actually a (real) homotopy invariant. At best, then, we could characterise the real homotopy type of the manifold. But even this is not the case: there are examples of manifolds of different real homotopy type that nevertheless have the same cohomology ring. A classic example, due to Borel, is the homogeneous space $Sp(5)/SU(5)$ vs. the connected sum $(S^6 \times S^{25})\#(S^{10} \times S^{21})$. These two spaces have the same cohomology ring, but nevertheless do not have the same homotopy type. Another example is the complement of three unlinked circles in $\mathbb{R}^3$ vs. the complement of the Borromean rings in $\mathbb{R}^3$. In the light of these examples, one would be tempted to say that cohomology cannot distinguish them, but this is not strictly true. The cohomology *as a commutative associative algebra* cannot distinguish them, but there exist higher algebraic structures which can and do distinguish them. These higher algebraic structures take the form of $n$-ary products called *Massey products*, which can in principle be computed from a knowledge of the de Rham complex, if not the de Rham cohomology ring. In fact, the algebraic structure in cohomology which derives from the Massey products is called a *commutative strongly homotopy associative algebra*.

Do these structures find their analogues in the context of topological conformal field theories? Presumably yes. Although no explicit examples have been constructed, there is little reason to assume that Massey products do not exist in the present context—after all, they do exist in Lie algebra cohomology and, in many cases, the BRST cohomology of a topological conformal field theory is the semi-infinite cohomology of some Lie algebra. In fact, the cohomology of a topological conformal algebra conjecturally admits the structure of a *homotopy BV algebra*; although a precise definition is lacking at present.

Taking this at face value, we would then say that the strict equivalence of topological conformal field theories consists of an isomorphism of the BRST cohomologies as homotopy BV algebras. This presumably coincides with the concept of *physically* equivalent TCFTs, which simply means that the correlation functions of the two theories should agree under the isomorphism. We can therefore refine our concept of cohomological equivalence to one of *homotopy equivalence*. That said, we believe that the method used in this paper to construct cohomological equivalences, would also construct homotopy equivalences, since the method involves tensoring with a Koszul TCA which is a "contractible" TCA. The results in this paper for the string theories and in [10] for the Kazama algebra, suggest the following improvement to the conjecture in [7]:

*Every TCFT is homotopy equivalent to a twisted $N{=}2$ SCFT.*

If true, the concept of a homotopy BV algebra is to be looked for no further than in the structure already present in the chiral ring of any twisted $N{=}2$ superconformal algebra. This would then hint strongly at the fact that the geometric origin of a homotopy BV algebra is to be found in the moduli space of $N{=}2$ super-Riemann surfaces.

Let us conclude with some remarks concerning the proof of the conjecture. It seems clear that in order to proceed further we have to identify the obstruction to the "$N{=}2$-ness" of a topological conformal algebra homologically. This is prompted by the fact that whatever this obstruction is, we are killing it by tossing in a Koszul TCA—which is very reminiscent of the Koszul–Tate procedure for killing (co)homological obstructions. In other words, we are in the embarrassing situation where we know how to kill the obstruction, yet we lack a clear understanding of what it is that we are killing. We believe that the homological interpretation of the obstruction to "$N{=}2$-ness" is the essential prerequisite to be able to formulate more precisely and eventually prove the above conjecture. In this regard the analogy with Liouville theory may play a role.

## ACKNOWLEDGEMENTS

It is a pleasure to thank Füsun Akman for some lively e-correspondence on cohomology BV algebras and for sending me her proof of (T4). I would like to thank Takashi Kimura for some early collaboration. This paper would not have been possible without our many conversations on this topic. It was our lack of success in finding a conformal algebra obeying (T1)–(T3) but not (T4) which originated the research reported here. In particular, the proof of (T4) at the end of Section 2 was obtained with him. It was also he who brought to my attention the topological analogy of the concluding section and the example of the Massey products on the complement of the Borromean rings. I am grateful to Jim Stasheff for the Borelian example of Massey products, for the term "cohomologically equivalent" and for his comments on a previous draft of this paper, as well as for sharing some of his immense knowledge and insight on things homotopic. Sasha Voronov was the first to point out to me that what I had called "equivalence" in [7] was actually weaker. It is a pleaasure to thank him for this and also for his insightful correspondence on topological conformal field theories. To all three collectively: Takashi Kimura, Jim Stasheff and Sasha Voronov, I owe many thanks for having taught me a great deal on this subject. I am also grateful to Jaume Roca for his comments on a previous draft of this paper.

Much of the work described in this paper was done while I was a guest of the Departments of Mathematics and Physics of the University of North Carolina at Chapel Hill. I would like to express my deep gratitude to Louise Dolan and Jim Stasheff for the initial invitation and for the warm hospitality with which I was received. Special thanks go also to Takashi Kimura for putting up with me for so long without complaining. Last but not least, I am sincerely grateful to Louise Dolan for her timely and decisive intervention to put a happy end to my (over)stay in Chapel Hill.